\documentclass[twocolumn,showpacs,preprintnumbers,amsmath,amssymb]{revtex4}

\usepackage{graphicx}
\usepackage{dcolumn}
\usepackage{bm}

\begin{document}


\title{Ionization of highly charged relativistic ions by neutral atoms
and ions}


\author{G.Baur}
\affiliation{Institut f\"ur Kernphysik and J\"ulich Centre
for Hadron Physics, Forschungszentrum J\"ulich, 
D-52425 J\"ulich, Germany}

\author{I.L. Beigman,  V.P. Shevelko, I.Yu. Tolstikhina}
\affiliation{P.N. Lebedev Physical Institute, Leninskii prospect
53, 119991 Moscow, Russia}

\author{Th. St\"ohlker}
\affiliation{GSI Helmholtzzentrum f\"ur Schwerionenforschung
GmbH, Planckstrasse 1, D-64291 Darmstadt, 
and Physikalisches Institut, Ruprecht-Karls-Universit\"at
Heidelberg, Philosophenweg 12, D-69120 Heidelberg, Germany}


\date{\today}

\begin{abstract}
Ionization of highly charged relativistic ions 
by neutral atoms and ions is considered. Numerical results
of recently developed computer codes based on the relativistic
Born and the equivalent-photon approximations are presented.
The ionization of the outer shells dominate.
For the outer projectile electron shells, which give the main contribution to the process, 
the non-relativistic Schr\"odinger wave functions can be used. 
The formulae for the non-relativistic reduction
of the Dirac matrix-elements are obtained for ionization of electrons 
with arbitrary quantum numbers $n$ and $\ell$. 
\end{abstract}


\maketitle


\section{Introduction}

 Projectile ionization, also referred to as electron loss or stripping,
 is an important charge-changing process playing a critical role in many
 applications such as heavy-ion driven inertial fusion, ion-beam lifetimes
 in accelerators, medical science, material  technology
and others (see, e.g., \cite{AtPhys} -- \cite{Kraft}).

In order to calculate the non-relativistic 
ionization cross sections for ion-atom collisions  
a computer code called LOSS was developed
using the Born approximation \cite{She} with account for the
screening as well as antiscreening effects. Non-relativistic
Schr\"odinger wave functions are used to describe the electronic
structure. This is a good approximation for the outer shell
electrons, which play the dominant role.

Relativistic ionization has been pioneered at Stanford by Anholt
and collaborators \cite{anh} and it is reviewed in several books
and review articles (see e.g., \cite{Ei}--
\cite{VoitkivPhysRep2004}). Numerical calculations of relativistic
ionization cross sections are presented in the literature mainly
for ionization of H- and He-like ions from the ground 1s state
(see, e.g., \cite{anh85} - \cite{Voitkiv2007}).

Recently, two new computer codes LOSS-R (Relativistic Loss) and HERION
(High Energy Relativistic IONization) have been
developed for calculation of the relativistic ionization cross
sections. The LOSS-R code \cite{LOSS-R} was created on the basis
of the non-relativistic LOSS code \cite{She} using the relativistic
Born approximation in the
momentum-transfer representation without magnetic interactions. 
The HERION code \cite{HERION}
uses the dipole and impulse approximations with relativistic wave
functions but it also neglects the magnetic interactions.

It is the aim of the present work to provide a theoretical
framework for ionization in relativistic ion-atom collisions
with possible account for the magnetic interactions 
for an arbitrary many-electron ions. The formulae obtained can be used for
projectile ions such as
U$^{28+}$ colliding with the rest-gas atoms and molecules at
energies up to a few GeV/u. Such heavy many-electron ions are of
practical implications for the international FAIR project at GSI
Darmstadt \cite{FAIR}.

In section 2 the main formulae for relativistic ionization are
recapitulated, also in order to establish the notation. Numerical
results due to the LOSS-R and HERION codes are given
in section 3. In section 4 the non-relativistic reduction of the
Dirac matrix-elements is done. 


\section{First order perturbation theory of relativistic
ionization}

 An accurate procedure to include the
relativistic effects in ionization cross sections using the plane
wave Born approximation is given in
\cite{Ei}. There, the
cross section summed over the magnetic quantum numbers and presented
in the momentum-transfer $q$-representation is given in
the form:
\begin{widetext}
\begin{equation}
\frac{d\sigma}{dE}=8\pi (Z_T\alpha)^2 \left(\frac{c}{v}\right)^2 \int_{q_0}^\infty
\frac{dq}{q^3}(|F(q)|^2 + \frac{\beta^2(1-q_0^2/q^2)}{(1-\beta^2
q_0^2/q^2)^2} |G(q)|^2), \label{basic}
\end{equation}
\end{widetext}
where $\alpha = e^2/\hbar c$ denotes the fine-structure
constant, $Z_T$ the effective charge of the target atom,  $q_0 =
\omega/v$, and $\omega$ is the ionization energy. The
 ion velocity is given by the relativistic factor
$\beta=v/c$. The matrix-elements $F$ and $G$ are evaluated in a x,
y, z coordinate system in which the z-axis lies along the momentum
transfer vector $\vec q$. The x-axis lies in the plane formed by
the vectors $\vec q$ and the projectile velocity $\vec v$. They
contribute incoherently to the cross section due to different
selection rules for the final magnetic substates. The matrix
elements  $F$ and $G$ are given by \cite{Ei}
\begin{equation}
F(q) \equiv <f|e^{iqz}|i>,
\label{F}
\end{equation}
\begin{equation}
G(q) \equiv <f|\alpha_x e^{i q z}|i>.
\label{G}
\end{equation}
where $|i>$ and $|f>$ denote the wave functions in the initial and final states and
$\alpha_x$ the x-component of the Dirac matrix vector $\vec{\alpha}$.
This is a useful splitting: the $F$-term tends to a constant for
$\gamma \rightarrow \infty$ whereas the $G$-term increases as
$ln\gamma$  due to a contribution of the 'equivalent' photons. The
matrix element $F$ is the main term in the non-relativistic
ionization, e.g. in the formulae of the LOSS code \cite{She}
whereas the matrix element $G$ appears for relativistic
ionization. The $G$-term can be calculated numerically if the
Dirac wave functions with the large and small components are
known.

The equivalent photon approximation can be obtained from eq. (1):
 Due to the photon pole,
the second term will dominate at high energies. For small $q$-values the quantity
$G(q)$ is directly related to the dipole matrix element:
\begin{equation}
G(0)= \frac{m \omega}{\hbar ^2}<f|z|i> \equiv
\frac{m \omega}{\hbar ^2} D_{fi}.
\end{equation}
The integral over $q$ in eq. (1) has a dominant contribution
at $q \sim q_0$. In the integration over $q$, we may assume that $G(q)=G(q_0)$.
Then the integral diverges logarithmically at large $q$. One can introduce a suitable
cutoff $q_{max}$ which corresponds to a cutoff  impact parameter $b_{min}$
in the impact-parameter space.
 In turn, the dipole moment
determines the photoionization cross section (see e.g., \cite{besa}, eq. (69.2)):
\begin{equation}
\sigma (\omega)= \frac{2\pi e^2 \hbar^2}{m^2 c \omega} |D_{fi}|^2
\end{equation}
So the ion-atom ionization cross section can be expressed via a photoionization
cross section and the equivalent-photon number.

The PWBA (Plane-Wave Born Approximation) formalism is also used in the calculation of bound-free
pair production in antiproton-nucleus
collisions (antihydrogen production) \cite{helm1}
and in  heavy ion collisions \cite{helm2}.
Pair production may be viewed as ionization of the negative
energy Dirac sea.
In the pair production
process one has to use Dirac wave functions, rather than the
Schr\"odinger wave functions.

In \cite{HERION} the dipole approximation was realized as a part
of  the HERION code where the dipole part of relativistic
ionization cross section of the projectile ion in collision with a
neutral atom is expressed via photoionization cross section
$\sigma(\omega)$ (see \cite{bertu}):
\begin{equation}
\sigma_{dip}(v) = \int_{\omega_{min}}^\infty n(\omega) \sigma
(\omega)\frac {d\omega}{\omega},
\end{equation}
where $\omega_{min}$ denotes the ionization threshold and
$n(\omega)$ the equivalent-photon number
\begin{widetext}
\begin{equation}
n(\omega) = \frac{2{Z_T}^2\alpha}{\pi} \left(\frac {c}{v}\right)^2
\left[x K_0(x) K_1(x)- \frac{1}{2}(\beta x)^2 (K_1^2(x)-K_0^2(x))
\right],\,\, x = \frac {\omega_{min}\, b_{min}}{v\gamma}.
\label{xx}
\end{equation}
\end{widetext}
where $K_n(x)$ is the McDonald function. The parameter $b_{min}$
is defined by the size of the projectile-ion shell $nl$ from
which ionization occurs and is given with a good accuracy by
\begin{equation}
b_{min} \approx \frac{n}{\sqrt{I_{nl}/Ry}}a_0,
\end{equation}
where $a_0$ = 0.53 $\times  10^{-8}$ cm is the Bohr radius,
$I_{nl}$ denotes the binding energy of the $nl$ shell and
$Ry$ = 13.606 eV is the Rydberg energy unit.
\section{Numerical calculations}

\begin{figure}
\begin{center}
\includegraphics[width=0.5\textwidth ]{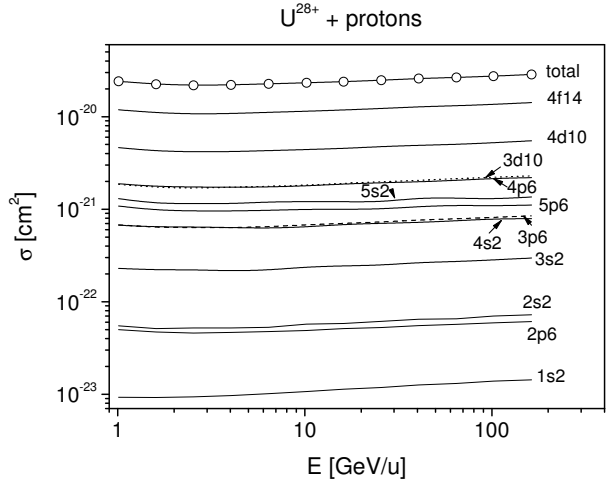}
\caption{Calculated ionization cross sections of
U$^{28+}(1s^2...4f^{14} 5s^2 5p^6)$ ions by protons: the LOSS-R -
code \cite{LOSS-R}. Contribution of ionization from different
subshells of U$^{28+}$ are shown together with the total cross
section.}
 \label{F:U28_proton}
\end{center}
\end{figure}

Calculated partial (on projectile electron subshells $nl$) and
total ionization cross sections of U$^{28+}$ ions by protons at
energies $E$ = 1 - 100 GeV/u are shown in Fig.~\ref{F:U28_proton}.
Calculations were performed by the LOSS-R code including dipole
and non-dipole parts of the ionization cross sections and using
the non-relativistic wave functions. As seen from the figure, the
main contribution to the total cross section is given by electron
ionization from 3$d^{10}, 4p^6, 4d^{10}$,..., shells which can be
described by the non-relativistic wave functions

\begin{figure}
\begin{center}
\includegraphics [width=0.5\textwidth]{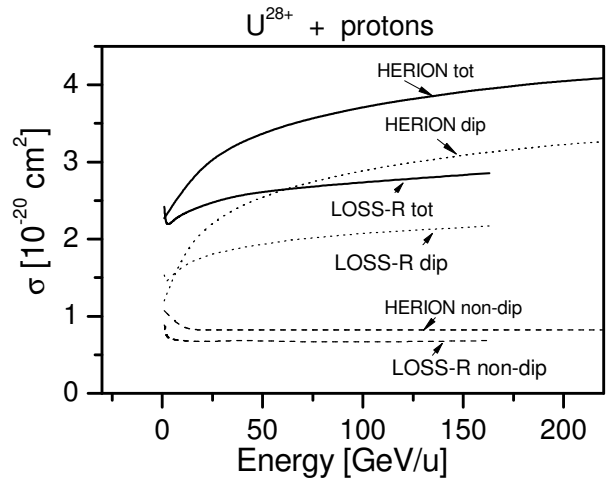}
\caption{
Relativistic dipole, non-dipole and total ionization
cross sections, i.e. summed over all $nl$ shells, of U$^{28+}$ by
proton impact calculated by the HERION and LOSS-R codes
(indicated): dashed curves - the non-dipole parts, dotted curves -
the dipole parts, and solid curves - the total cross sections. }
 \label{F:U28prot}
\end{center}
\end{figure}

A comparison of relativistic ionization cross sections of
U$^{28+}$ ions by proton impact calculated by two different codes, LOSS-R and
HERION, is shown in Fig.~\ref{F:U28prot} where the contributions from dipole-,
non-dipole parts are shown together with the total (dipole + non-dipole) cross sections.
In the LOSS-R code, the dipole part of the cross sections
corresponds to electron transitions into the continuum with the
orbital angular momentum quantum numbers 
$\lambda = l \pm 1$ where $l$ is the
angular momentum of the projectile-electron shell. As seen from the
figure, the non-dipole parts calculated by the codes agree within
30 \% meanwhile the dipole parts and the total cross sections
agree only within a factor of 2. Most probably, this discrepancy
is related to two reasons: first, the use of different wave
functions in the codes, i.e., relativistic wave functions in HERION
code and and non-relativistic ones in the LOSS-R code. And second,
in the HERION code dipole ionization cross sections are calculated
on the basis of photoionization cross sections; the latter are
much higher than those calculated in the usual non-relativistic
approximation because of the presence of the so-called giant
resonances occurring  due to configuration interaction of the
final states (see \cite{Amusya} for details).

We note that the dipole part of ionization cross section involves
about 60-70 \% of the total cross section while the relativistic
non-dipole part has a weak dependence on energy and, therefore,
the dipole part has practically the same shape as the total cross
section.

\begin{figure}
\begin{center}
\includegraphics [width=0.5\textwidth] {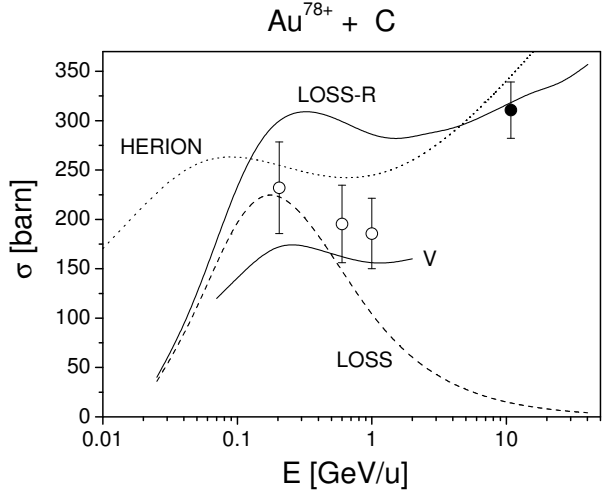}
\caption{Ionization cross sections of H-like Au$^{78+}(1s)$ ions
by carbon atoms. Experiment: open circles - \cite{Tomas97},
\cite{Tomas94} solid circle - \cite{Claytor}. Theory: LOSS -
non-relativistic LOSS code \cite{She}, LOSS-R - relativistic
LOSS-R code \cite{LOSS-R}, HERION - HERION code \cite{HERION} with
effective target charge $Z_T^2 = Z^2 + N $ = 42, V - calculations
by Voitkiv \cite{Voitkiv2007} with relativistic electron
description.}
 \label{F:Au78C}
\end{center}
\end{figure}

Relativistic ionization cross sections of Au$^{78+}$(1s) ions in
collisions with carbon are displayed in Fig.~\ref{F:Au78C} where
experimental data are given by symbols and theoretical
calculations by curves. The non-relativistic result (LOSS)
shows a Born maximum followed by a decrease of the cross section.
The relativistic result (LOSS-R) has a local minimum around 1
GeV/u energy and increases logarithmically with energy. The
results obtained with the HERION code have a little better agreement with
experiment at $E >$ 200 MeV/u than those by the LOSS-R code. In
the calculations, the effective charge of the target $Z_T = Z^2 +
N$ = 42 was used while in the LOSS-R code the $Z_T$ value is
found in the $q$-representation with account of all subshells of
the target (see eq. 27 and Ref. \cite{She} for details). The lower curve is the
recent result \cite{Voitkiv2007} with relativistic electron
description.

\section{Non-relativistic limit of
$\alpha$-matrix element}
Now we consider the non-relativistic limit of the
matrix-element G given by eq. (\ref{G}):
\begin{equation}
G(q)=<f|\alpha_x exp{(i q z)}|i>. \nonumber\\
\end{equation}
The operator $J=\alpha_x \exp{(i q z)}$ is an odd one and connects
the large and small components of the Dirac wave function. It
gives zero if one only uses the large components. The
non-relativistic limit of $G(q)$ is given in \cite{mes} in the
form:
\begin{equation}
J=\frac{1}{2m}(\alpha_x exp{(i q z)} \rho_1 (\vec \sigma \cdot
\vec p) + (\vec \sigma \cdot \vec p)\rho_1 \alpha_x \exp{(i q z)}),
\label{J}
\end{equation}
where $\vec p =\vec e_z q$. This operator acts only on the large
component of the Dirac wave function. We use the relations $\alpha_x \rho_1
=\sigma_x$ and $\rho _1 \alpha_x=\sigma _x$, (see \cite{mes}, p. 891) to obtain
\begin{equation}
J=\frac{1}{2mc}(\sigma_x \exp{(i q z)} (\vec \sigma \cdot \vec p)+
(\vec \sigma \cdot \vec p) \sigma _x \exp{(i q z)}).
\end{equation}
Using the identity $(\vec \sigma \cdot \vec A) (\vec \sigma \cdot
\vec B)=(\vec A \cdot \vec B)+ i \vec \sigma \cdot (\vec A \times
\vec B)$ (see e.g. eq.  XIII.83 of \cite{mes}) with $\vec A = \vec
e_x; \vec B = \vec p$ and  with $\vec B = \vec
e_x; \vec A = \vec p$ we find
\begin{equation}
J=\frac{1}{2m}( \exp{(i q z)} p_x + p_x \exp{(i q z)} + 
\rm{spin-flip}\, terms). \label{J1}
\end{equation}
The spin-flip terms are given by
\begin{equation}
\exp{(i q z)} (\vec p \times \vec \sigma)_x -(\vec p \times \vec
\sigma)_x \exp{(i q z)}
\end{equation}
In the following, we neglect the spin-flip terms because they are
small in our case, see also \cite{anh}). So the non-relativistic 
limit of the odd operator $J$ is
\begin{equation}
J=\frac{1}{2m}(\exp{(iqz)}p_x + p_x \exp{(iqz)}. \label{J2}
\end{equation}
This result can be also obtained 
directly using the electromagnetic-current approach in the Schr\"odinger
theory (see, e.g., \cite{landau}, Chapter $82a$).

The spin-flip terms appear in the Pauli approximation,  which can
be obtained from the Dirac equation. The matrix element $G(q)$
becomes
\begin{equation}
G(q)= \frac{-i\hbar}{mc} <f|\exp(iqz)\frac{\partial}{\partial
x}|i>.
\end{equation}

Here $p_x=-i\hbar \frac{\partial}{\partial x}$ and
$\frac{\partial}{\partial x} e^{iqz}=0 $.

The derivative $\frac{\partial}{\partial
z}$ is easier to calculate in spherical coordinates rather than 
$\frac{\partial}{\partial x}$,
so we rotate the coordinate system by $\pi/2$ around the y-axis.
This changes $ x\rightarrow z$ and $z\rightarrow -x$. So we have
\begin{equation}
G(q)= \frac{-i\hbar}{mc} <f|\exp{(-iqx)}\frac{\partial}{\partial
z}|i>.
\end{equation}
Actually we only need the absolute square of the function $G$, summed over spins and
integrated over the angle of the outgoing electron. This matrix
element is calculated according to the rules of angular momentum
algebra (see e.g., \cite{edm}). The bound state wave function $|i>$
is presented in the form $|i>=f_{l_i}(r)Y_{l_im_i}(\hat r)$, i.e. with separated angular
and radial parts.

The exponential factor is expanded into partial waves in the usual
way:
\begin{equation}
\exp{(-iqx)}=4 \pi \sum_{lm} i^{+l}j_l(qr)
Y_{lm}(\hat{r})Y^*_{lm}(\hat{x})
\end{equation}

The radial wave function of the bound and continuum states satisfy
the normalization conditions:
\begin{widetext}
\begin{eqnarray}
\int_0^\infty g^2_{l_i}(r) dr = 1, \,\,\,\,\,\label{norma decrete}
g_{l_f}(r) \approx  \frac{1}{\sqrt k} \,{\rm sin}\left(kr +
\frac{1}{k} {\rm ln}(2kr) + \eta \right), \,\,\,k^2/2 =\epsilon,
\label{norma continuum}
\end{eqnarray}
\end{widetext}
where $\epsilon$ is the energy of ejected electron and $\eta$ is
the scattering phase shift.


According to \cite{edm} , the action of $\partial/\partial z$ on
this wave function yields:
\begin{widetext}
\begin{eqnarray}
\partial \Psi/\partial z =
\left(\frac{l+1}{\sqrt{(2l+1)(2l+3)}} Y_{l+1 m} +
\frac{l}{\sqrt{(2l-1)(2l+1)}} Y_{l-1 m}\right)\partial
f_l/\partial r
\nonumber \\
-(\frac{l(l+1)}{\sqrt{(2l+1)(2l+3)}}Y_{l+1m}-\frac{l(l-1)}{\sqrt{
(2l-1)(2l+1)}}Y_{l-1m})f_l/r.
\end{eqnarray}
\end{widetext}
Thus, the matrix element is presented as a sum of angular and
radial parts. We note that the present procedure is quite similar
to that used for calculating the magnetic terms of the pionium
breakup \cite{taheim}.

We define two types of radial matrix elements, a 'usual' one
\begin{equation}
R^B_{lf\lambda li}(q)\equiv \int_0^\infty dr g_{lf}
j_\lambda(qr)f_{li},
\label{RB}
\end{equation}
 and a 'new' one involving the derivative of the initial radial wave function:
\begin{equation}
R^d_{lf\lambda li}(q(q)\equiv \int_0^\infty dr r g_{lf} j_\lambda
(qr) \frac{df_{li}}{dr}.
\label{Rd}
\end{equation}
The three-dimensional integration is, as usual, decomposed into a
radial and angular parts. The angular integration can be done using
eq. (4.6.3) in \cite{edm} and $ Y_{lm}^*=(-1)^m
Y_{l-m}$.

In the integration over the three spherical harmonics one obtains
two types of terms: $\Omega_+$ and $\Omega_-$ with
\begin{widetext}
\begin{equation}
\Omega_\pm = \sqrt{\frac{(2l_f +1)(2 \lambda + 1)(2l_i \pm 1)}
{4\pi}} \left(
\begin{array}{ccc}
l_f & \lambda & l_i \pm 1\\
m_f & \mu & m_i\\
\end{array}
\right) \left(
\begin{array}{ccc}
l_f & \lambda & l_i \pm 1 \\
0 & 0 & 0\\.
\end{array}
\right)
\end{equation}
\end{widetext}
Collecting all the factors one obtains
\begin{widetext}
\begin{eqnarray}
&G&=\sum_{lf mf \lambda \mu} \frac{4\pi^2}{k}\, i^\lambda \,
Y^*_{\lambda \mu} (\hat x) Y^* _{l_f m_f}(\hat
k)\nonumber\\&\times& \left(\frac{l_i+1}{\sqrt{(2l_i + 1)(2l_i +
3)}} \Omega_+ (R^d - l_i R)  + \frac{l_i}{\sqrt{(2l_i -1)(2l_i
+1)}} \Omega_-(R^d -(l_i -1)R)\right)
\end{eqnarray}
\end{widetext}
As was mentioned above, we need  the function $|G(q)^2|$
integrated over $\Omega_k$ and summed over spins.
The integration over $\Omega_k$ makes the sum over $l_f$ and $m_f$
incoherent, due to the orthogonality of the spherical harmonics.
We
assume that the states  $j=l\pm 1/2$ are
degenerate. Using the sum rules, the summation over $m_i$ and $m_f$ can be
easily done:
\begin{widetext}
\begin{eqnarray}
\sum_{m_i m_f} \left(
\begin{array}{ccc}
l_f & \lambda & l_i \pm 1\\
m_f & \mu & m_i\\
\end{array}
\right) \left(
\begin{array}{ccc}
l_f & \lambda^\prime & l_i \pm 1 \\
m_f & \mu^\prime & m_i\\
\end{array}
\right) =\delta_{\lambda, \lambda^\prime} \delta_{\mu, \mu^\prime}
/(2 \lambda+1)
\end{eqnarray}
\end{widetext}

The summation over $\mu$ can be also done using the completeness relation
\begin{equation}
\frac{4\pi}{2\lambda + 1}\sum_\mu Y^*_{\lambda \mu} (\hat x)
Y_{\lambda \mu}(\hat x)=1,
\label{sum_rule}
\end{equation}
and one obtains  the function $|G(q))|^2$  for given projectile 
quantum numbers $n_i, l_i$ 
\begin{widetext}
\begin{eqnarray}
|G(q)|^2=\frac{2\hbar^2}{(mc)^2} \sum_{\lambda \mu l_f} |(l_i+1)
u_+ (R^d -l_i R)+ l_i u_-(R^d+(l_i -1)R|^2
\end{eqnarray}
\end{widetext}
The radial integrals $R^B$ and
$R^d$ depend on $q$, $l_f, \lambda,\, l_i$ according to eqs. (19) and (20), 
and $u_\pm$ is given by
\begin{eqnarray}
u_\pm =\frac{\pi}{k} i^{\lambda}
\sqrt{\frac{(2l_f+1)(2\lambda + 1}{2l_i+1}} \left(
\begin{array}{ccc}
l_f & \lambda & l_i \pm 1 \\
0 & 0 & 0\\.
\end{array}
\right)
\end{eqnarray}

We note that $\lambda=0$ corresponds to
the dipole excitation, and a monopole contribution does not exist for
the matrix element $G$ but corresponds to the 'equivalent photon' contribution
with the photon having spin 1. The
monopole part is included to the the matrix-element $F$ (see eqs. (1) and (2)).

Finally, for relativistic ionization cross section from the projectile electronic
shell by a heavy target particle one can use eq. (1) 
but replacing the target effective charge
$Z_T$ by the $q$-dependent charge $Z_T(q)$ in the form:
\begin{widetext}
\begin{equation} Z_T^2(q)=\left(Z-\sum_{j=1}^{N_T}
F_{jj}(q)^2 \right)^2+\left(N_T-\sum_{j=1}^{N_T}
F_{jj}(q)^2\right),\,\,\,
 F_{JJ}(q)=<j|{\rm exp}{(i{\bf qr})}|j> ,
\label{Rtarget}
\end{equation}
\end{widetext}
where $Z$ and $N_T$ denote the nuclear charge and the number of electrons of the target
($Z = N_T$ for neutral atoms and $Z$ = 1, $N_T$ = 0 for protons).
The form of eq. \ref{Rtarget} makes it possible
to take into account the screening and anti-screening effects of all
target electrons.

The function $F(q)$ in eq. (1) is the 'usual' Born matrix element which after separating
angular and radial parts is written in the form:
\begin{equation}
|F(q)^2| = (2\lambda +1)(2l_f+1) \left(
\begin{array}{ccc}
l_i & l_f & \lambda \\
0   &   0  &  0 \\
\end{array}
 \right)^2 |R^B(q)|^2,
\label{Born}
\end{equation}
where the Born integral $R^B(q)$ is given in eq. (\ref{RB}).

The function $G(q)$ describes the magnetic interactions between projectile ion and target atom
and has the form:
\begin{widetext}
\begin{eqnarray}
|G(q)|^2 & = &\frac{2}{c^2}\frac{(2\lambda+1)(2l_f+1)}{2l_i+1}
\nonumber\\
&\times & \left | i^{\lambda} \left(
\begin{array}{ccc}
l_f & \lambda & \l_i+1\\
0 & 0 & 0\\
\end{array}
\right)  \lbrack (l_i+1)R^d(q)+l_i(l_i-1)R^B(q)  + i^\lambda
\left(
\begin{array}{ccc}
l_f & \lambda & \l_i-1\\
0 & 0 & 0\\
\end{array}
\right) \lbrack l_i R^d(q) +l_i(l_i-1)R^B(q)\rbrack \right |^2 ,
\label{derivative}
\end{eqnarray}
\end{widetext}
where the radial integral $R^d(q)$ is given by eq. (\ref{Rd}).

\section{Conclusion}
Relativistic ionization of heavy ions colliding with atoms or ions
is considered in the plane-wave Born approximation (PWBA). The
ionization cross sections are presented in the momentum-transfer
$q$-representation as a sum of two terms: the 'usual' Born
approximation (matrix element of $i\bf{qr}$) and the
relativistic term responsible for the magnetic interactions
between colliding particles and expressed through the x-component
$\alpha_x$ of the Dirac matrix vector $\vec{\alpha}$.

A simple limit of PWBA is the equivalent photon approximation
(dipole approximation). Numerical calculations  for the dipole and
non-dipole parts of relativistic cross sections are presented for
the p + U$^{28+}$ and C+Au$^{78+}$ collisions using the LOSS-R  and
HERION computer codes.

A new formula is obtained for 
the relativistic  ionization cross section
for an arbitrary $nl$ shell of the projectile ion in the form of
separated angular and radial parts. The radial part is 
expressed via an integral of the initial bound state wave
function and the derivative of the final continuum wave function. 
In this way, new possibilies to investigate 
ionization processes including many electron atoms 
are opened.

\section{Acknowledgments}
The authors are grateful to S.N. Andreev for his help in 
the codes development. This work was performed under the INTAS grant Nr.
06-1000012-8530 and RFBR grant Nr. 08-02-00005-a.

\end{document}